\documentclass[prb,twocolumn,showpacs,psfig,superscriptaddress,longbibliography]{revtex4-2}
\usepackage{amssymb}
\usepackage{algpseudocode}
\usepackage{algorithm}
\usepackage{amsmath}
\usepackage{braket}
\usepackage{booktabs}
\usepackage{chngcntr}
\usepackage{color}
\usepackage{comment}
\usepackage{dsfont}
\usepackage{etoolbox}
\usepackage{epstopdf}
\usepackage[T1]{fontenc}
\usepackage{float}
\usepackage{graphicx}
\usepackage[colorlinks=true,linkcolor=blue,citecolor=blue,urlcolor=blue]{hyperref}
\usepackage[latin9]{inputenc}
\usepackage{indentfirst}
\usepackage{latexsym,bm,euscript}
\usepackage{mathrsfs}
\usepackage{multirow}
\usepackage{natbib}
\usepackage{soul}
\usepackage{subfigure}
\usepackage{times}
\usepackage{tikz}
\usepackage{txfonts}
\usepackage{ulem}
\usepackage{xspace}
\usepackage{xfrac}
\usepackage{physics}

\def\para{\ensuremath{/\kern -0.8em /}\xspace}
\def\ket#1{|{#1}\rangle}                 

\def\beqn{\begin{eqnarray}}
\def\eeqn{\end{eqnarray}}
\def\beq{\begin{equation}}
\def\eeq{\end{equation}}

\newcommand{\Beq}{\begin{eqnarray*} }
\newcommand{\Eeq}{\end{eqnarray*} }
\newcommand{\Bmat}{\left(\begin{matrix}}
\newcommand{\Emat}{\end{matrix}\right)}

\newcommand{\bbE}[1]{\mathbb{E}\left[#1\right]}
\newcommand{\dP}[0]{\mathrm{d}\mathbb{P}}

\begin{document}


\title{Finite-Temperature Simulations of Quantum Lattice Models with Stochastic Matrix Product States} 

\author{Jianxin Gao}
\thanks{These authors contributed equally to this work.}
\affiliation{Peng Huanwu Collaborative Center for Research and Education, 
School of Physics, Beihang University, Beijing 100191, China}
\affiliation{CAS Key Laboratory of Theoretical Physics, Institute of 
Theoretical Physics, Chinese Academy of Sciences, Beijing 100190, China}

\author{Yuan Gao}
\thanks{These authors contributed equally to this work.}
\affiliation{Peng Huanwu Collaborative Center for Research and Education, 
School of Physics, Beihang University, Beijing 100191, China}
\affiliation{CAS Key Laboratory of Theoretical Physics, Institute of 
Theoretical Physics, Chinese Academy of Sciences, Beijing 100190, China}

\author{Qiaoyi Li}
\email{liqiaoyi@itp.ac.cn}
\affiliation{CAS Key Laboratory of Theoretical Physics, Institute of 
Theoretical Physics, Chinese Academy of Sciences, Beijing 100190, China}
\affiliation{School of Physical Sciences, University of Chinese Academy of Sciences, Beijing 100049, China}

\author{Wei Li}
\email{w.li@itp.ac.cn}
\affiliation{CAS Key Laboratory of Theoretical Physics, Institute of 
Theoretical Physics, Chinese Academy of Sciences, Beijing 100190, China}
\affiliation{Peng Huanwu Collaborative Center for Research and Education, 
School of Physics, Beihang University, Beijing 100191, China}

\date{\today}
\begin{abstract}
In this work, we develop a stochastic matrix product state (stoMPS) 
approach that combines the MPS technique and Monte Carlo 
samplings and can be applied to simulate quantum lattice models 
down to low temperature. In particular, we exploit a procedure to 
unbiasedly sample the local tensors in the matrix product states, 
which has one physical index of dimension $d$ and two geometric 
indices of dimension $D$,
and find the results can be continuously improved by enlarging $D$. 
We benchmark the methods on small system sizes and then 
compare the results to those obtained with minimally entangled 
typical thermal states, finding that stoMPS has overall better 
performance with finite $D$. We further exploit the MPS sampling
to simulate long spin chains, as well as the triangular and square 
lattices with cylinder circumference $W$ up to 4. 
Our results showcase the accuracy and effectiveness of stochastic 
tensor networks in finite-temperature simulations.
\end{abstract}
\maketitle

\section{Introduction}
Finite-temperature calculations of quantum many-body system play an 
indispensable role in the studies of quantum matter and materials. It bridges 
the gap between quantum lattice models and experiments in a wide range 
of investigations, ranging from studies of highly frustrated quantum magnets, 
unconventional superconductivity, to the ultracold atom quantum simulations. 
In frustrated quantum magnets, the finite-temperature 
approach can help determine the microscopic spin models from fitting the 
measured thermodynamic properties~\cite{Chen2019,Li2020TMGO,Li2021NC,
Gao2022NBCP,YuCPL2021}, including the specific heat, magnetic susceptibility, 
and also spin dynamics at finite temperature, providing insight into the 
quantum spin states in the compounds~\cite{Chen2019,Li2020TMGO,
Li2021NC,HLi2020PRR,Larrea2021Nature,Wang2023PRLPlaquette}.
It can also be exploited to study the exotic low-temperature electron states 
in the fermion Hubbard model~\cite{Wietek2021PRX,Chen2022tbg,Qu2022tJ,
Qu2023bilayer}, enabling an unbiased and accurate comparison with optical 
lattice quantum simulations~\cite{Mazurenko2017,Koepsell2021,Chen2021SLU}.

Tensor networks offer a feasible method for partially overcoming the 
exponential wall in quantum many-body simulations. Beyond the 
ground-state properties, various thermal tensor-network algorithms 
were proposed for accurate finite-$T$ calculations~\cite{Bursill1996DMRG,
Wang1997,Xiang1998Thermodynamics,Zwolak2004,Feiguin2005,
White2009METTS,Stoudenmire2010,Li2011,Czarnik2012PEPS,
METTSvsPurification2015,Dong2017,Chen2017,Chen2018,Li2019,
tanTRG2023}. Currently, the finite-$T$ tensor-network methods can 
be classified into two major categories, purification and typical thermal 
states. The former exploits tensor-network representations, e.g., matrix 
product operator (MPO) and projected entangled pair operator (PEPO), 
of the thermal density matrix, and can be used to simulate both 1D and 
2D systems~\cite{Bursill1996DMRG,Wang1997,Xiang1998Thermodynamics,
Feiguin2005,Zwolak2004,Feiguin2005,Li2011,Czarnik2012PEPS,Dong2017,
Chen2017,Chen2018,Li2019,tanTRG2023}. The latter, with a representative 
method called minimally entangled typical thermal states (METTS)
\cite{White2009,Stoudenmire2010,Bruognolo2015,Goto2020,Wietek2021PRX,
Wietek2021PRXMott}, constructs a Markov chain samplings of matrix
product state (MPS)~\cite{MPS1992,MPS1993} with very short 
self-correlation length.

\begin{figure}[htp]
\includegraphics[width=1\linewidth]{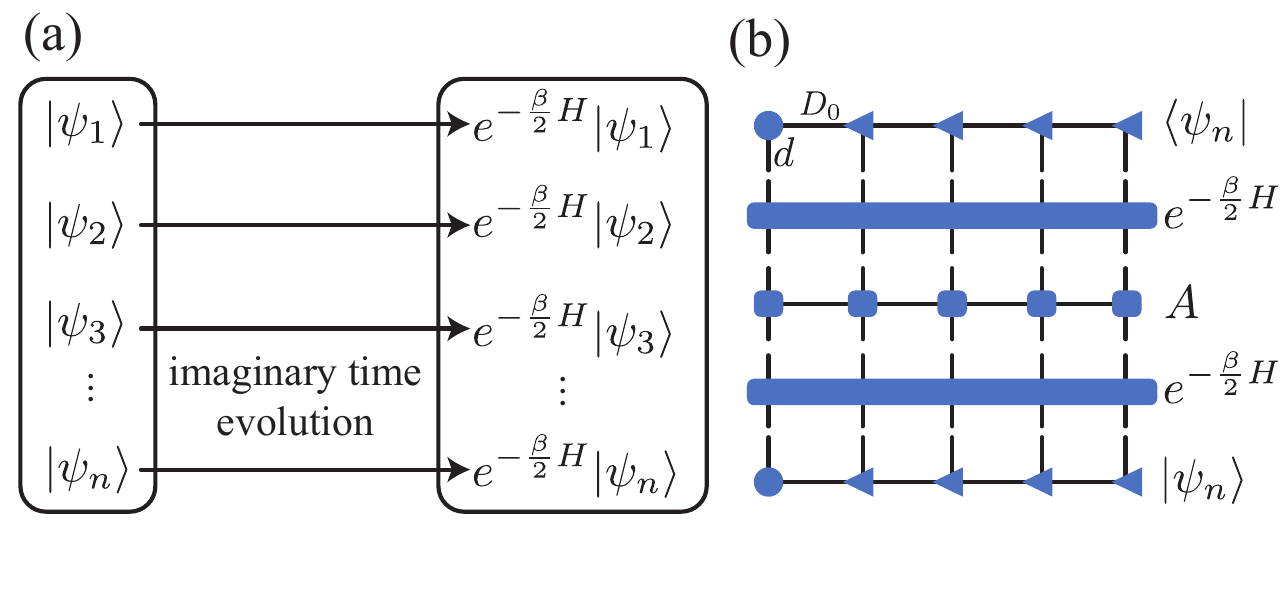}
\caption{In the stoMPS algorithm, we (a) sample the isometric MPS
$\ket{\psi_n}$ in a given sample space, and (b) calculate the expectation 
values of quantity $A$ on the time-evolved MPS $e^{-\frac{\beta}{2} H} 
\ket{\psi_n}$.}
\label{Fig:Fig1}
\end{figure}

Both approaches have their own pros and cons. The MPO-based 
approaches benefits from the high precision and controllability 
in 1D and quasi-1D systems with certain widths $W$. The local 
tensor in MPO have two physical indices, different from that of 
the MPS with a single physical index. At sufficiently low temperature,
the geometric bond dimension of MPO is also believed to be 
significantly larger than that in the ground-state MPS. Therefore, 
it is a nice idea switching from MPO to MPS to improve the efficiency 
in the low-temperature limit~\cite{White2009,Stoudenmire2010}. 
However, in practical calculations of both 1D and 2D lattice systems, 
it has been demonstrated that MPS-based Monte Carlo sampling, 
such as that used in METTS, is still less efficient and less accurate 
than the MPO-based approach~\cite{METTSvsPurification2015,
Chen2018,Li2019,tanTRG2023}. Therefore, there is still a great 
need to develop more efficient stochastic MPS methods to further 
enhance the performance of this hybrid approach that combines 
tensor networks and Monte Carlo sampling.

In this work, we introduce a highly efficient stochastic matrix product state 
(stoMPS) approach inspired by the finite-temperature Lanczos method (FTLM)
\cite{Jaklic1994,Huang2018,Schnack2020}, which is used for evaluating 
systems of small sizes. FTLM combines the Lanczos diagonalization technique 
with random sampling and converting the problem of diagonalizing the 
Hamiltonian in the full Hilbert space to the Krylov subspace generated 
from some random initial states. We device the stoMPS algorithm as a 
generalization of FTLM and make detailed comparisons between results 
obtained with several different sample spaces. We find that sampling in 
a continuous sample space shows better performance, even outperforming 
the METTS method with Markov-chain sampling. Furthermore, we 
demonstrate the high scalability of this approach by applying it on 2D 
cylinders with width up to $W = 4$. The connections of our stoMPS 
approach to the thermal pure quantum (TPQ) states~\cite{Garnerone2010,
Sugiura2012,Sugiura2013,Iwaki2021} are also discussed. 

The rest part of the article is arranged as follows. In Sec.~\ref{Sec:Alg} we 
introduce the stoMPS algorithm and compare it to FTLM as well as METTS
methods. The applications of stoMPS approach to 1D spin chains, square,
and triangular lattice Heisenberg models are presented in Sec.~\ref{Sec:App},
and Sec.~\ref{Sec:Con} is devoted to the summary.

\section{Stochastic Matrix Product State Algorithm}
\label{Sec:Alg}

\subsection{Sampling Algorithm with Matrix Product States}
The stoMPS workflow is depicted in Fig.~\ref{Fig:Fig1}, 
which is a tensor-network generalization of FTLM to large 
system size. In the FTLM method (see Appendix~\ref{App:FTLM}), 
a random vector is uniformly selected from the unit sphere of 
the many-body Hilbert space $\mathcal{H}$ as the starting point, 
and a Krylov space is constructed based on this initial state. 
The total Hamiltonian is then projected into this subspace using 
the Lanczos technique, and the average value is computed. 
By repeating this process, one can obtain the finite-temperature 
properties of many-body systems, despite limitations in system 
size, with high accuracy~\cite{Jaklic1994,Huang2018,Schnack2020}.

To extend this approach to larger systems, we resort to stochastic 
MPS states instead of random initial vectors. A crucial question that 
needs to be addressed for this generalization is how to perform MPS 
samplings in a manner that represents the unit sphere unbiasedly. 
It is notable that with any given sample space $\{\ket{\psi}\} 
\subset{\mathcal{H}}$, we have 
\begin{equation}
A(\beta) := \frac{{\rm \tr}[e^{-\beta H} A ]}{{\rm \tr}[e^{-\beta H}]} = 
\frac{\bbE{\bra{\psi} e^{-\beta H} A \ket{\psi}} }{ \bbE{\bra{\psi} e^{-\beta H} \ket{\psi}}}, 
\end{equation}
for an observable $A$ if and only if a proper probability is chosen,
such that the expectation satisfies
\begin{equation}
\bbE{\ket{\psi}\bra{\psi}} \propto I, 
\label{Eq:unbais}
\end{equation}
where $\beta$ is the inverse temperature, and $I$ is the identity in 
the Hilbert space. 

One way to make $\ket{\psi}$ satisfy condition Eq.~(\ref{Eq:unbais}) 
is to restrict the sample space to the subset of all direct product states 
($D=1$ MPS), which simplifies the condition to a local version, i.e., 
$\bbE{\ket{s_i}\bra{s_i}} \propto I_{\mathcal{H}_i}$ for each site $i$, where 
$\mathcal{H}_i$ is the local Hilbert space and $I_{\mathcal{H}_i}$ is 
the identity operator. Specifically, we can let $\ket{s_i}$ distribute 
uniformly on $\{\uparrow, \downarrow\}$ (dubbed as $Z_2$ sampling 
hereafter) or on the unit sphere [dubbed as U(1) sampling], which is 
equivalent to the $D = 1$ case of the random isometry sampling 
introduced below for general $D \geq 1$ cases), i.e., $\ket{s_i} = 
\cos \theta \ket{\uparrow} + \sin \theta \ket{\downarrow}$, where 
$\theta \sim {\rm U}[0, 2\pi]$ for spin-1/2 systems with local Hilbert 
space dimension $d=2$.

\begin{figure}[htp]
\includegraphics[width=1\linewidth]{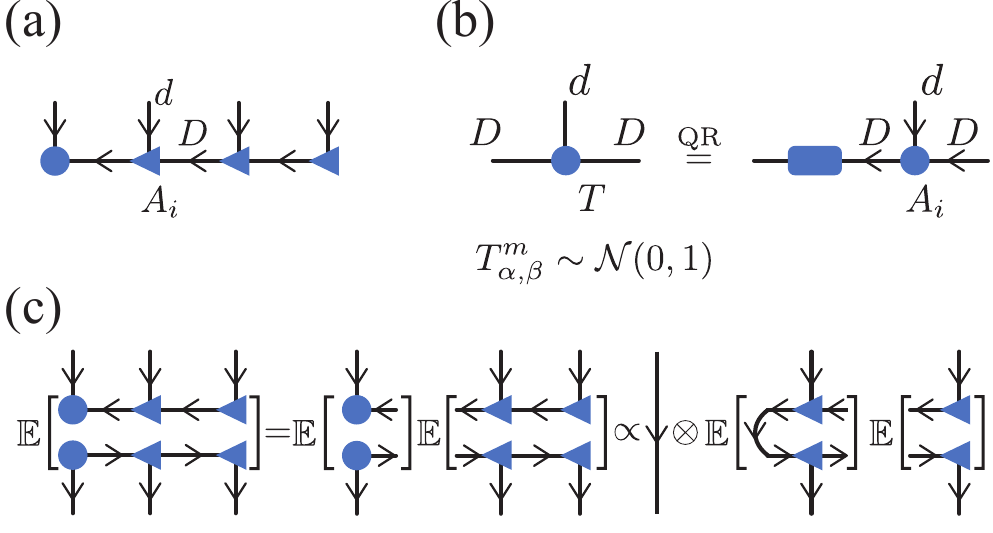}
\caption{(a) The construction of stochastic isometric MPS, which 
satisfies the canonical condition following the order specified by 
the arrows. (b) To sample the isometric MPS, we conduct QR 
decomposition of the random matrix $T_{\alpha, \beta}^m$ whose 
elements are independently generated according to the standard 
normal distribution $\mathcal{N}(0,1)$. 
(c) The expectation of $\ket{\psi}\bra{\psi}$ in MPS representation, 
which can be reduced sequentially into product of identity, and thus 
satisfy Eq.~(\ref{Eq:unbais}).
}
\label{Fig:Fig2_Proof}
\end{figure}

Note the physical meaning of the isometric local tensors of a canonical 
MPS can be understood as the sequentially selected  renormalization 
basis. Inspired by this, we can naturally generalize the sampling from 
$Z_2$ and U(1) to the $A$ tensor in MPS with finite $D$. We sample 
random isometries and require the final center tensor distribute uniformly 
on the unit sphere of local renormalized Hilbert space. We find it is 
sufficient to satisfy Eq.~(\ref{Eq:unbais}) if the isometries distributed 
on Stiefel manifold St($D, Dd$) according to the Haar measure 
\footnote{More strictly speaking, the induced measure is a homogeneous 
space of O($Dd$).} and are independent amongst different sites. 
The expectation $\bbE{\ket{\psi}\bra{\psi}}$ can be decomposed 
from site to site, then one can find the total tensor network represents 
an identity via recursively using a lemma on random isometry
(see Appendix~\ref{App:Lemma}) from the left to the right, see Fig.~\ref{Fig:Fig2_Proof}(c). 

To be practical, the random isometries are generated via QR decomposition 
of a $D\times d \times D$ random tensor where each element is generated independently according to standard normal distribution $\mathcal{N}(0,1)$
\cite{FASI2021297,Birkhoff1979}, c.f. Fig.~\ref{Fig:Fig2_Proof}(b). Besides
the sampling scheme shown in Fig.~\ref{Fig:Fig2_Proof}, there exists an
alternative approach to obtain the initial MPS from random unitary MPO 
[see Appendix~\ref{App:MPO} and also Ref.~\cite{Garnerone2010}].

\begin{figure}[htp]
\includegraphics[width=0.85\linewidth]{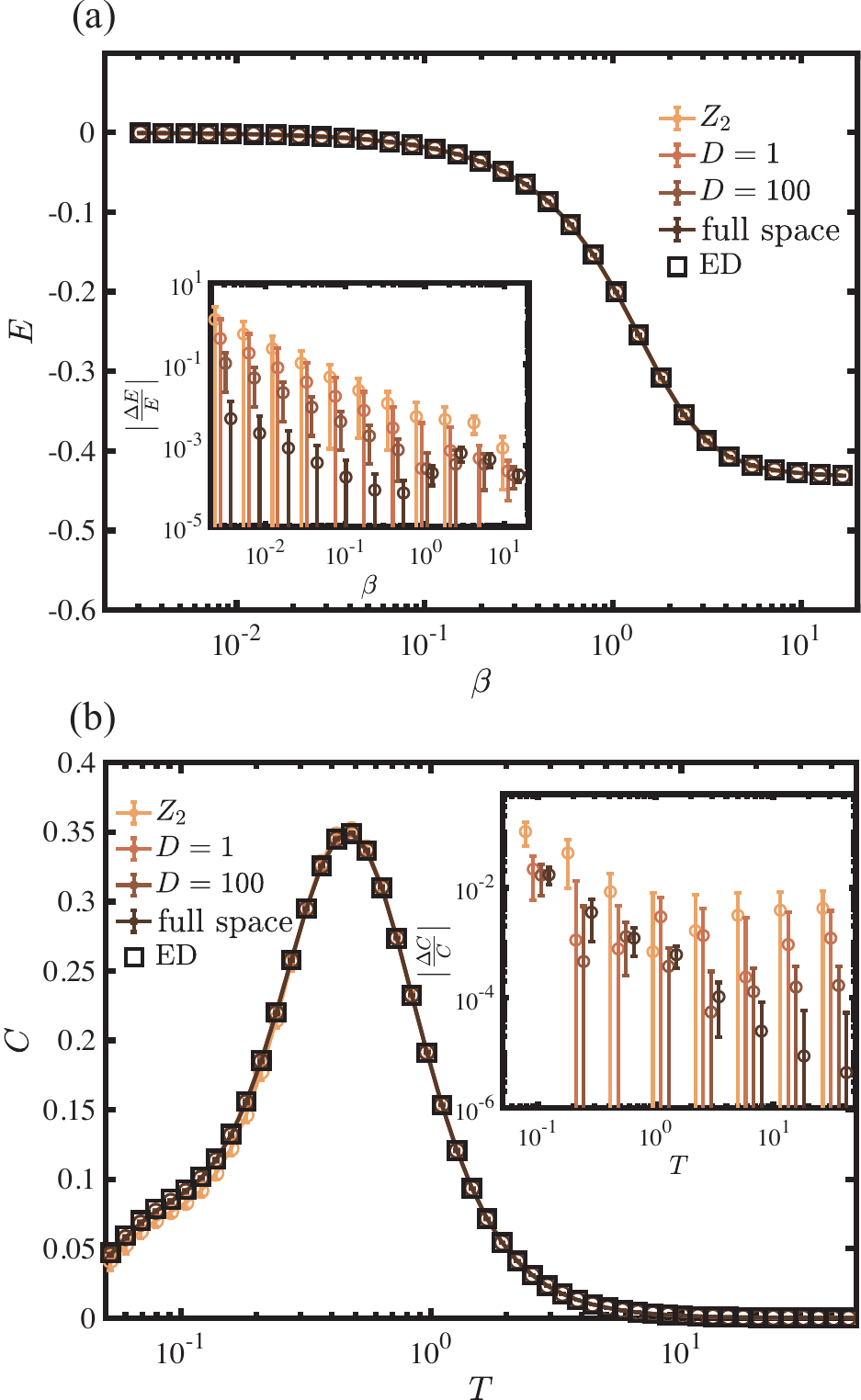}
\caption{The energy and specific heat results of the $L = 16$ 
Heisenberg chain. (a) shows the the energy results, where various 
sampling schemes generate results in excellent agreement with ED. 
The inset shows the statistical errors, which are very small and 
continuously improved as the initial bond dimension $D$ increases, 
approaching the results of FTLM with equivalently full rank $D$. 
(b) shows the specific heat results, where the inset shows that the 
standard errors also decrease with $D$.}
\label{Fig:Fig2} 
\end{figure}

\subsection{Benchmark on the Heisenberg Spin Chain}
Below we showcase the accuracy and efficiency of the stoMPS approach 
with different sampling schemes on a $L=16$ Heisenberg chain with XXZ
Hamiltonian $$H = \sum_i J_{xy} (S_i^x S_{i+1}^x + S_i^y S_{i+1}^y) 
+ J_z S_i^zS_{i+1}^z.$$ The calculated energy and heat capacity with 
corresponding stand errors are shown in Fig.~\ref{Fig:Fig2}. 
At low temperatures, the $Z_2$ sampling performs poorly, 
as the overlap between the initial state and the ground state vanishes 
when $\langle S^z_{\rm tot} \rangle \neq 0$ initial state is selected
randomly. However, as $D$ increases we find the mean value approaches 
the ED results, with stand errors of energy expectation $\sigma[E]$ 
and heat capacity $\sigma[C]$ also decrease, as shown in the insets 
of Fig.~\ref{Fig:Fig2}. We also find the standard errors in stoMPS 
with large $D$ converge to that obtained by FTLM. The latter samples 
vectors in the Hilbert space instead of in the MPS space, and is thus 
equivalently a full-ranked MPS that can represent the Hilbert space globally. 

\begin{figure}[htp]
\includegraphics[width=1\linewidth]{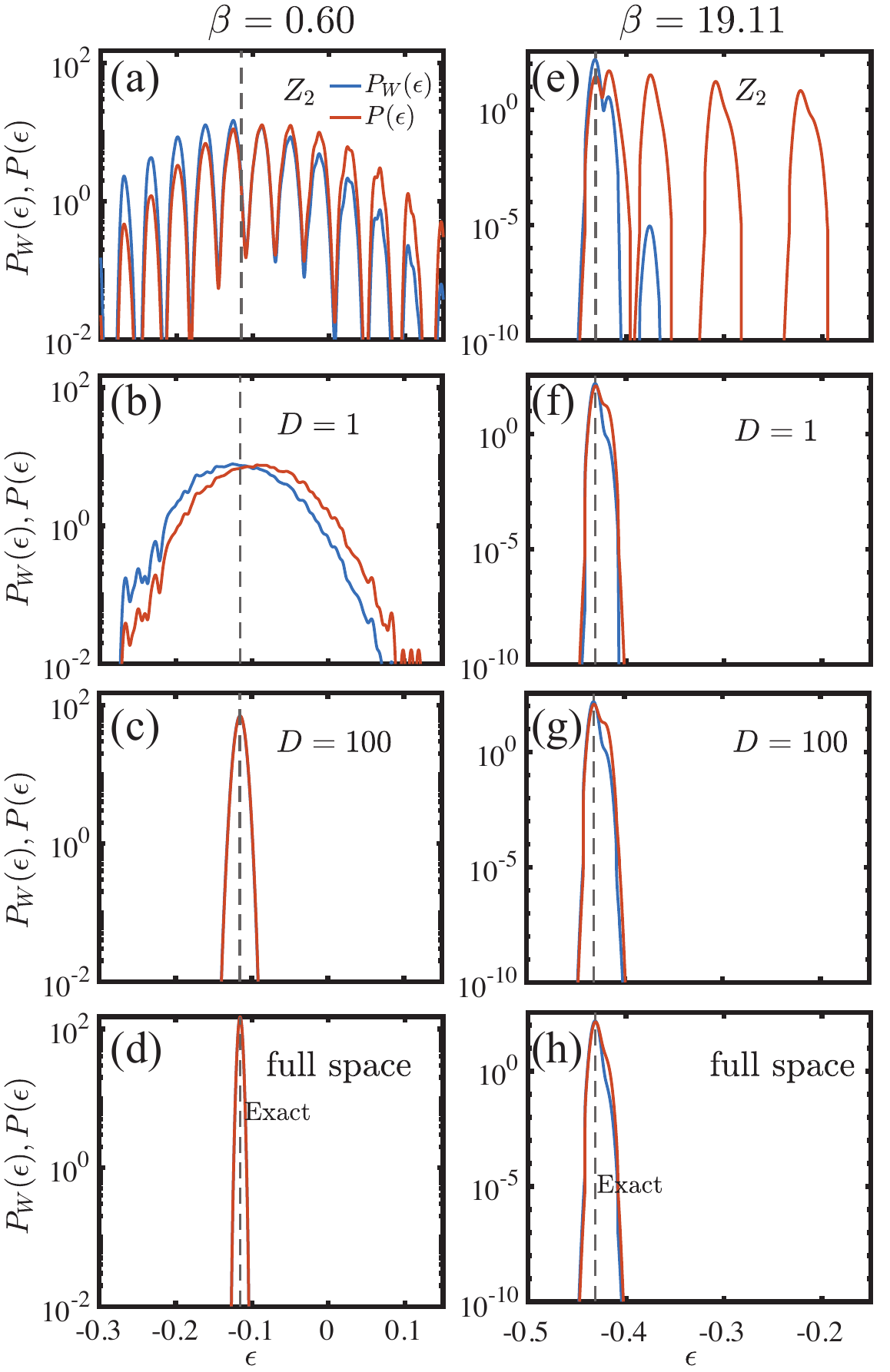}
\caption{The probability distribution $P(\epsilon)$ and weighted 
$P_{W}(\epsilon)$ at high $\beta = 0.60$ (left, a-d) and low 
temperature $\beta = 19.11$ (right, e-h) obtained with different 
sample spaces: $Z_2$, $D = 1$ [U(1)], $D = 100$ and full sample 
space. In the calculations, we evolved the states $\ket{\Psi}$ 
including finite-$D$ MPS and vector.}
\label{Fig:Fig3} 
\end{figure}

\begin{figure}[h]
\includegraphics[width=1\linewidth]{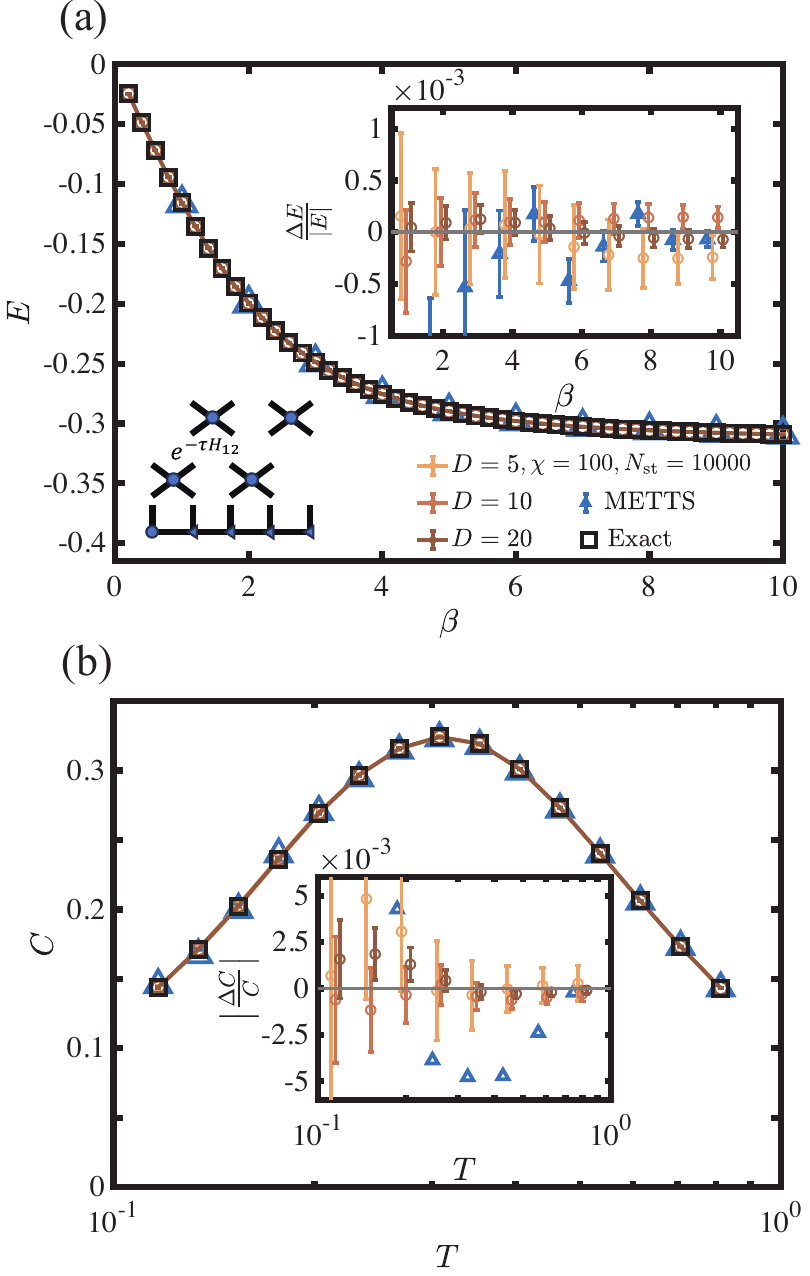}
\caption{The results of energy and specific heat of the $L = 50$ XY chain. 
(a) The energy results are well converged by retaining $\chi=100$ bond 
states and with $N_{\rm s}=10,000$ samples. Increasing bond dimension 
$D$ of the initial MPS can continuously improve the accuracy, as shown 
in the upper right inset of (a). The METTS results with $\chi = 100$ and 
$N_{\rm s} = 10,000$ are shown as a comparison. Left bottom inset 
illustrates the TEBD technique employed in the calculations. (b) The 
stoMPS results of specific heat show high accuracy when compared 
to the analytical solution, which are better than the METTS results 
shown in the inset. Reliable error bars for METTS results of specific 
heat is not available due to tricky issues in numerical differentiation.}
\label{L50_E_C}
\end{figure}

To further analyze the sampling efficiency, we now move from sample space 
to energy space and estimate the unweighted and weighted probability 
density from Monto Carlo samplings 
\begin{equation}
	P(\epsilon) \simeq \frac{1}{N_{\rm s}} 
\sum_n K(\epsilon-\epsilon_n; \sigma)
\label{Eq:P_e}
\end{equation}
and
\begin{equation}
P_{W}(\epsilon) \simeq \frac{\sum_n \expectationvalue{e^{-\beta H}}{\psi_n} 
K(\epsilon-\epsilon_n; \sigma)}{\sum_n \expectationvalue{e^{-\beta H}}{\psi_n}},
\label{Eq:PW_e}
\end{equation}
where $N_{\rm s}$ is the sample size, $\epsilon_n$ is the energy expectation
value of the time-evolved sample $e^{-\beta H/2}\ket{\psi_n}$ and $K(\epsilon; 
\sigma)$ is a gaussian kernel. The weighted distribution density can be related
with the energy expectation through (see Appendix \ref{App:C}) 
$$E= \int \epsilon P_W(\epsilon)d\epsilon.$$
In practice, we take $\sigma = 2.5\times10^{-3}$ and show the results 
with different sampling strategies in Fig.~\ref{Fig:Fig3}. 

In Fig.~\ref{Fig:Fig3}(a-d), we present the results at relatively high 
temperature, where the probability distribution $P_{W}(\epsilon)$ 
becomes sharper and approaches the exact results as the bond 
dimension $D$ of the sampled MPS increases, indicating that the 
sampling efficiency is enhanced. Specifically, we find that by increasing 
the sampling space, the distribution $P_{W}(\epsilon)$ becomes more 
concentrated, as shown in Figs.~\ref{Fig:Fig3}(a-d). At low temperature, 
i.e., Fig.~\ref{Fig:Fig3}(e-h), the discrete $Z_2$ sampling scheme 
is far less efficient compared to the continuous $U(1)$ sampling 
scheme ($D=1$). This can be ascribed to the large number of 
low-weight samples in the $Z_2$-sampling strategy. When the total 
spin $\langle S^z_{\rm tot} \rangle$ of the initial state is nonzero, 
the overlap between the initial state and the ground state vanishes. 
As shown in Fig.~\ref{Fig:Fig3}(e), only about 33\% of the samples have 
weights $W_n > 10^{-4}$. As shown in Figs.~\ref{Fig:Fig3}(f-h), when 
we increase the initial bond dimension $D$ and thus more random 
parameters, the sample distribution becomes more concentrated, 
with the sample weight also increased.

\section{Applications on large-scale quantum lattice models}
\label{Sec:App}

With the unbiased sampling schemes constructed, in the construction of 
Krylov subspace we need to carry out imaginary time evolution on the 
sampled initial MPS $\ket{\psi_n}$, i.e., $\ket{\beta_n} = e^{-\beta H/2}
\ket{\psi_n}$. We randomly generate certain MPS with a bond dimension 
of $D$ and conduct imaginary-time evolution to obtain $\ket{\beta_n}$, 
with which the thermodynamic observables can be computed (c.f., Fig.
\ref{Fig:Fig1}). In the course of imaginary-time evolution, bond dimension 
of the MPS will increase and a truncation of geometric bond is thus required. 
Here we employ the time-evolving block decimation (TEBD) technique
\cite{Vidal2004,Daley2004} for MPS and retain maximally $\chi$ bond states 
in the calculations. Below, we present results of stoMPS applied to long 1D 
spin chains, as well as 2D square- and triangular-lattice Heisenberg models 
on cylinders of finite widths.

\subsection{1D XY spin chain}
Now we consider a more realistic but still exactly soluable problem, 
a $L = 50$ XY chain, and showcase the powerfulness of stoMPS by 
calculating this model. To be specific, for the 1D Heisenberg chain 
$H = \sum_i h_{i,i+1}$, we exploit the TEBD technique to conduct 
the imaginary-time evolution on the MPS, which follows  
\begin{equation}
e^{-\beta H} = (e^{-\tau H})^N = 
(e^{-\tau H_{\rm even}} e^{-\tau H_{\rm odd}})^N + {O}(\tau),
\end{equation}
where $N \tau = \beta$, $H_{\rm even} = \sum_i h_{2i, 2i +1}$ and 
$H_{\rm odd} = \sum_i h_{2i-1, 2i}$. The stoMPS calculations are 
conducted with different bond dimensions $D$, all shared the same 
time evolution step length $\tau = 0.05$, and with a maximal bond 
dimension $\chi=100$. The results are averaged over $N_{\rm s} = 
10,000$ samples to obtain well converged results. 

In Fig.~\ref{L50_E_C}(a), we show the results of energy density and 
its relative errors with various initial $D$. The squares mark the 
analytical results of XY chain, and we find the stoMPS results, as 
well as METTS data, are in excellent agreement with the exact results. 
To see their relative errors, we show in the inset of Fig.~\ref{L50_E_C}(a)
$\Delta E/|E|$, and find the accuracy gets continuously improved as 
$D$ increases. At high to intermediate temperatures, e.g.,$\beta 
\lesssim 5$, stoMPS with even small $D$ clearly outperform METTS; 
at sufficiently low temperature, e.g., $\beta = 10$, the $D=20$ 
($\chi=100$) stoMPS data have very similar accuracy as compared 
to the METTS results. With the same truncated bond dimension 
$\chi=100$ and sample number $N_{\rm s}=10,000$, in practice
the METTS consumes 5 times CPU hours as compared to our stoMPS
method, making latter superior performance in versatility, accuracy,
and efficiency.

In Fig.~\ref{L50_E_C}(b), we present the specific heat results obtained 
using various methods. The black line represents the analytical solution, 
while the specific heat results computed by stoMPS were obtained through 
numerical differentiation. As shown in the figure, the results obtained by 
different methods are consistent with the analytical solution within the margin 
of error. The inset of Fig.~\ref{L50_E_C}(b) confirms that stoMPS has a clear 
advantage in computing the specific heat. stoMPS generates smoother data 
at different temperature points as it computes specific heat based on the 
same set of time-evolved MPSs, while METTS has to resample the procedure 
for each temperature point.

\subsection{Square and triangular-lattice spin models}
\begin{figure}[h]
\includegraphics[width=1\linewidth]{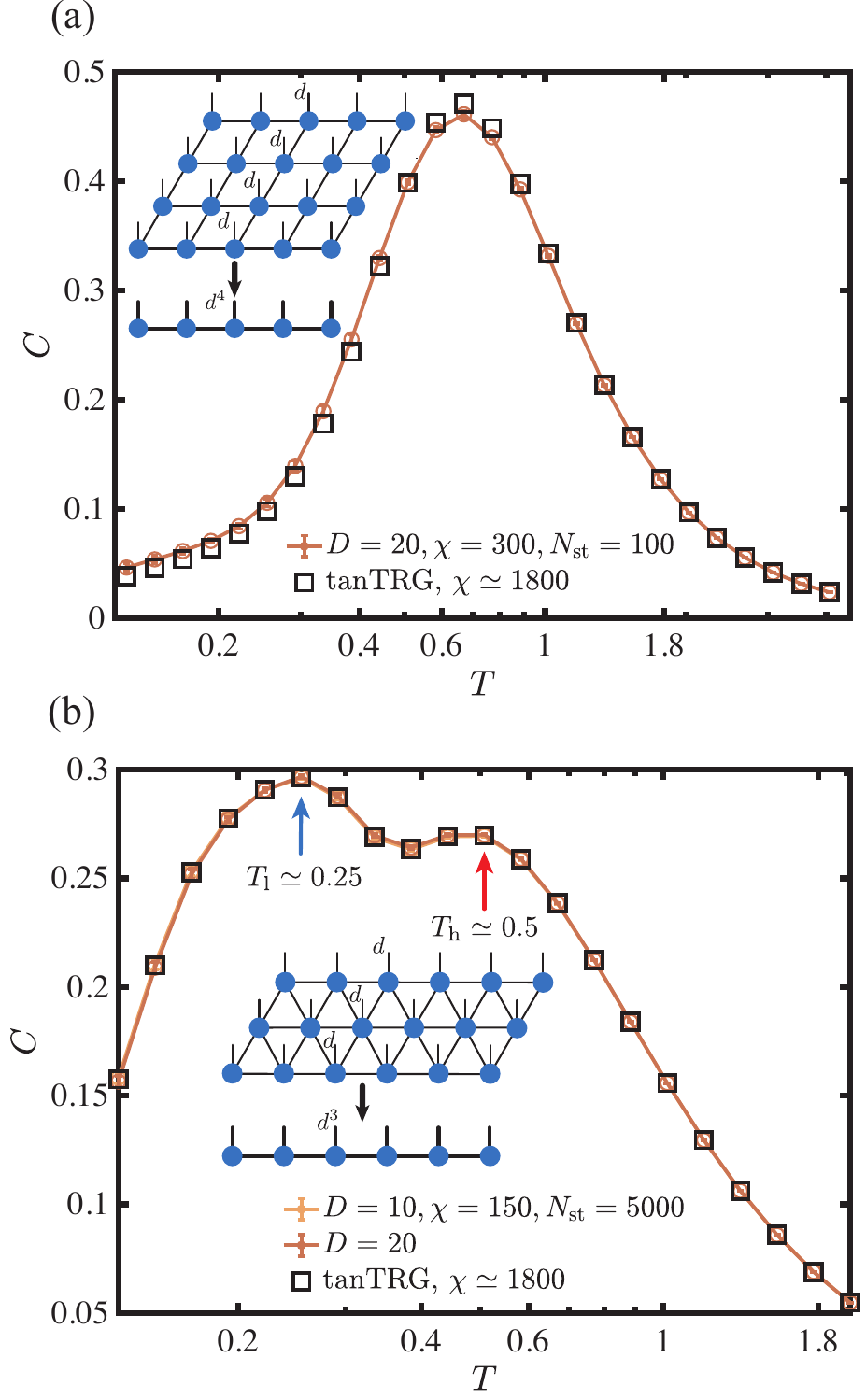}
\caption{(a) Specific heat results of Heisenberg model on $4\times8$ 
square lattice defined on cylinder geometry, and the stoMPS results 
with different $D$ are compared to that of tanTRG. 
(b) Results obtained on the $3\times6$ triangular lattice, which exhibits 
double-peak structure in sharp contrast to that of the square-lattice 
Heisenberg model in (a).}
\label{C_2D}
\end{figure}

Beyond 1D system, we employ stoMPS method to simulate 2D Heisenberg
antiferromagnetic lattice model wrapped on cylinder geometries. A conventional 
way to conduct such mapping is to follow the way routinely used in 2D density 
matrix renormalization group method. Here we showcase that the stoMPS can 
also be used to simulate the cylinders by ``compressing'' them into a 1D chain, 
as illustrated in the insets of Figs.~\ref{C_2D}(a,b). The corresponding MPS has 
a physical index with enlarged Hilbert space of dimension $d^W$, and the 
Hamiltonian only contains interactions between these nearest-neighboring 
composite sites. Thus, we can exploit TEBD techniques to simulate such 
systems similarly as in 1D chains, and compute the thermodynamics with 
high precision.

In Fig.~\ref{C_2D}(a), we present the results of the specific heat 
computed on a square lattice of size $4 \times 8$ (cylinder width 
$W=4$ and length $L=8$), where the results are in good agreement 
with the recent tangent-space tensor renormalization group (tanTRG)
approach, which is a state-of-the-art MPO-based method 
\cite{tanTRG2023} for many-body systems. The specific heat 
curve displays a bell-like shape, with a maximum located at 
approximately $T/J \approx 0.6$, where $J=1$ is the spin 
exchange. 

In addition to the unfrustrated spin model, stoMPS can also be used 
to investigate frustrated quantum antiferromagnets. In Fig.~\ref{C_2D}(b), 
we present the specific heat results computed on a $3\times 6$ cylinder. 
Previous numerical studies of triangular lattice Heisenberg antiferromagnets 
have revealed the presence of two 
specific heat peaks at $T_{\rm h}/J \sim 0.5$ and $T_{\rm l}/J \sim 0.2$,
respectively~\cite{Chen2018,Chen2019}, as also observed in experiments
\cite{Rawl2017,Cui2018}. It has been proposed that in the intermediate 
regime between the low-temperature scale $T_{\rm l}$ and the higher 
one $T_{\rm h}$, rotonlike excitations are activated with a strong chiral 
component and a significant contribution to the thermal entropy
\cite{Elstner1993,Elstner1994}. These gapped roton excitations
\cite{Zheng-2005,Zheng-2006}, which bear a striking resemblance 
to the renowned roton thermodynamics in liquid helium, suppress 
the incipient $120^\circ$ order that emerges for temperatures below 
$T_{\rm l}$.

Here in Fig.~\ref{C_2D}(b), even on a width-3 cylinder such a double-peak 
specific heat curve can be clearly identified, and the results are well 
converged by retaining only $D=10, 20$ bond states in the initialization 
and $\chi=150$ states during the imaginary-time evolution by TEBD. 
Our simulations on the square and triangular lattices show that the 
stoMPS method constitutes a practical approach for finite-temperature 
calculations of quantum lattice models, providing a valuable tool for
studying frustrated quantum magnetism.

\section{Summary and outlook}
\label{Sec:Con}
We construct an efficient algorithm for finite-temperature calculations
by combining tensor networks and Monte Carlo sampling. From sampled
MPS tensor $A$ with bond dimension $D$, we perform imaginary time
evolution and then obtain very accurate results over sample average.
We apply this method to spin chain and 2D spin systems including the 
Heisenberg model on the $4 \times 8$ square and $3 \times 6$ triangular 
lattices, and find the results are very accurate, where the sampling efficiency 
and accuracy can be improved by increasing the value of initial bond 
dimension $D$. Notably, we obtained two peaks of specific heat for the 
triangular lattice antiferromagnet. The algorithm has a number of advantages
over the MPO-based algorithm, including good parallelism and high efficiency 
for low-temperature simulations, and has an overall superior performance 
than the existing MPS-based Monte Carlo method METTS.

Although stoMPS exhibits promising performance, there are still several 
points that require further improvements. For instance, stoMPS is not an 
importance sampling technique and may require enhancements to improve 
its sampling efficiency. Additionally, for sufficiently wide 2D lattice systems, 
the projected entangled-pair state (PEPS) method may outperform the MPS 
method. Our work provides a foundation for generalizing stochastic tensor 
networks from MPS to PEPS and may lead to even more accurate and efficient 
simulation methods for 2D quantum lattice models at finite temperature. 

\section{Acknowledgement}
This work was supported by the National Natural Science Foundation of 
China (Grant Nos.~12222412, 11974036, and 12047503), and the CAS 
Project for Young Scientists in Basic Research (YSBR-057). We thank 
the HPC-ITP for the technical support and generous allocation of CPU time.

\appendix

\section{Finite-temperature Lanczos method}
\label{App:FTLM}
We briefly review the finite-temperature lanczos method (FTLM). 
Given a temperature $\beta = 1/T$, the measurement of an operator $A$ 
reads 
\begin{equation}
\langle A \rangle = \frac{1}{Z} {\rm Tr} [Ae^{-\beta H}]
= \frac{1}{Z} \sum_n \langle n |A e^{-\beta H} |n \rangle,
\end{equation}
where $\{|n\rangle\}$ represent an orthonormal basis of the Hilbert space and 
$Z = {\rm Tr}[e^{-\beta H}]$ is the partition function. 
Since the dimension of the Hilbert space increases exponentially 
as the system size increases, fully tracing becomes numerically impossible.
On the other hand, with a given sample space, we 
have 
\begin{equation}
	\langle A \rangle  = \overline{\langle A \rangle} := \frac{\bbE{\expectationvalue{Ae^{-\beta H}}{\psi}}} 
	{\bbE{\expectationvalue{e^{-\beta H}}{\psi}}},
	\label{Eq:Sample}
\end{equation}
if and only if 
\begin{equation}
	\bbE{\ket{\psi}\bra{\psi}} \propto I_N,
	\label{Eq:Unbiased condition}
\end{equation}
with $I_N$ the identity operator in the N-dimension Hilbert space.
Thus the measurement of $A$ can be obtained form a Monte Carlo
sampling process 
\begin{equation}
	\langle A \rangle  \simeq \frac{\sum_n^{N_{\rm s}}
\langle \psi_n | e^{-\beta H} A |\psi_n \rangle} {\sum_n^{N_{\rm s}} 
\langle \psi_n | e^{-\beta H}|\psi_n \rangle}, 
\label{Eq:Sample}
\end{equation}
where $N_{\rm s}$ is the sample size. In FTLM, the samples 
distribute uniformly on the unit sphere of the Hilbert space.

It remains to conduct the imaginary-time evolution of a given state 
using the Lanczos method, i.e., 
\begin{equation}
\begin{split}
e^{-\beta H}&= \sum_k^{\infty} \frac{-\beta^k}{k!} 
H^k\\
 & \simeq \sum_k^{\infty} \sum_r^{K} 
\frac{-\beta^k}{k!}  \epsilon_r^k |\psi_n^r\rangle \langle \psi_n^r | \\
&= \sum_r^{K} e^{-\beta \epsilon_r} 
|\psi_n^r\rangle \langle \psi_n^r |,
\end{split}
\end{equation}
where $K$ represents the dimension of the Krylov subspace generated by 
$\{|\psi_n\rangle, H|\psi_n\rangle, ..., H^{K-1} |\psi_n\rangle\}$ and
$|\psi_n^r\rangle$ is the $r$-th eigenvector of $H$ with energy $\epsilon_r$
in the Krylov subspace. 

The FTLM is a powerful tool that enables many-body calculations of the 
calculations of the finite-temperature and dynamic properties on finite-size
systems. However, the vector representation of many-body state has a
limitation as the computational cost grows exponentially with the system 
size. To extend the calculations to larger system sizes, a more efficient 
representation format such as the matrix product state (MPS) is required 
and developed in the present work.

\section{Lemma on random isometry}
\label{App:Lemma}
If $A \in {\rm St}(m \leq n,n)$ is a random isometry according to the Haar measure, then 
\begin{equation}
	\bbE{A^\dagger A} \propto I_{n\times n}.
\end{equation}
Proof: Let $O\in {\rm O}(n)$ be any orthogonal matrix, then 
\begin{equation}
	O^\dagger\bbE{ A^\dagger A}O = \bbE{O^\dagger A^\dagger A O} = \bbE{A^\dagger A} 
\end{equation}
since the Haar measure is invariant under the action of O($n$). The only matrix which 
permutes with the total O$(n)$ is the identity up to a coefficient, thus we have $\bbE{A^\dagger A} \propto I_{n\times n}$.

\section{Unitary MPO strategy}
\label{App:MPO}

\begin{figure}[htp]
\includegraphics[width=1\linewidth]{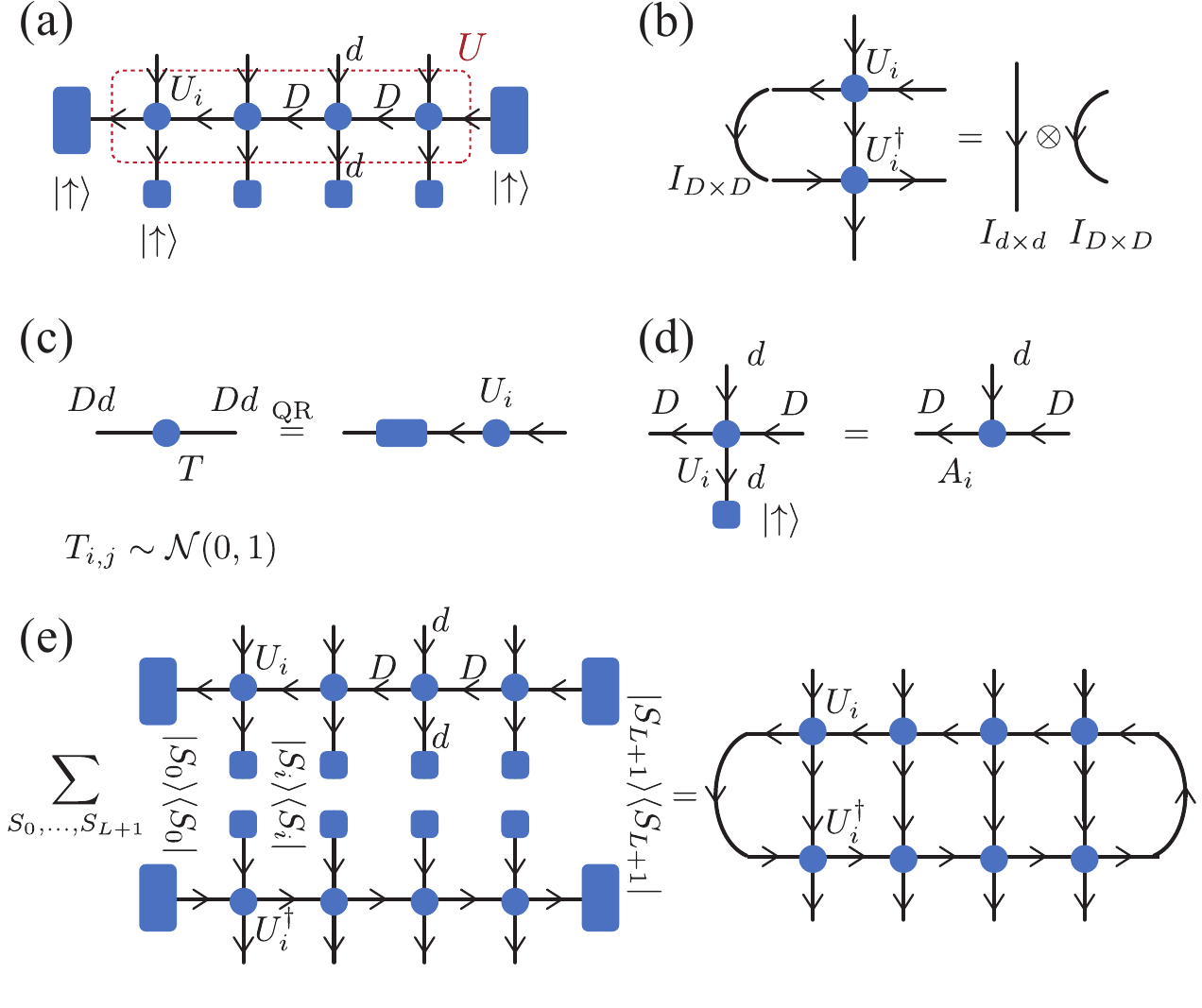}
\caption{(a) Generating a random MPS from unitary MPO. 
(b) The canonical condition of random unitary MPO. (c) To sample 
the unitary matrix, we conduct a QR decomposition of the random 
matrix $T_{i,j}$ where each element is generated according to the 
standard normal distribution $\mathcal{N}(0,1)$. (d) The isometric 
tensor can be obtained form the unitary matrix with $\ket{\uparrow} 
:= (1,0,...,0)^{\rm T}$. (e) Tensor-network illustration of Eq.~(\ref{Eq:B1}).}
\label{Fig:Fig7_Proof_P}
\end{figure}

In this section, we will introduce an alternative method to obtain a random 
MPS satisfying Eq.~(\ref{Eq:unbais}). Note that the sample spaces of either
$Z_2$ or U$(1)$ scheme introduced in the main text can be generated 
by a group of unitary operators acting on a trivial ferromagnetic (FM) 
state. Specifically, the $Z_2$ sampling corresponds to spin flipping 
($\mathbb{Z}_2^L$), while the U$(1)$ sampling corresponds spin 
rotation $\prod_{1}^L$U(1).
Based on this observation, we can construct a more general sample 
space by generalizing the unitary operation from local spin flip or 
rotation to a composite operation represented by a unitary matrix 
product operator (MPO) of finite bond dimension $D>1$, which applies
to the FM state and generate a stochastic initial MPS [see 
Fig.~\ref{Fig:Fig7_Proof_P}(a)].

Here, we demonstrate that the random MPS $\ket{\psi}$ we obtain 
satisfies condition~(\ref{Eq:unbais}). To obtain the random isometric 
tensor, we proceed as follows. First, we generate a random $Dd 
\times Dd$ unitary matrix according to the Haar measure, as shown in Fig.~\ref{Fig:Fig7_Proof_P}(c,d). Then, we select the first $D$ rows 
of this matrix to construct an $D \times Dd$ isometric matrix. Additionally, 
we introduce two $D$-dimensional auxiliary states $\ket{\uparrow}$ 
and $\ket{\uparrow}$ at the boundaries to eliminate redundancy.

Note that a local spin-flip action on site $i$ (or boundary state) can 
be absorbed into the $U_i$ without changing the probability, 
i.e, $P(U\ket{\uparrow,\uparrow,...,\uparrow} 
\bra{\uparrow,\uparrow,...,\uparrow}U^\dagger) 
= P(U\ket{\uparrow,\downarrow,...,\uparrow} 
\bra{\uparrow,\downarrow,...,\uparrow}U^\dagger)$. 
As shown in Fig.~\ref{Fig:Fig7_Proof_P}(e), we have
\begin{equation}
\begin{split}
&\bbE{U\ket{\uparrow,\uparrow,...,\uparrow}
\bra{\uparrow,\uparrow,...,\uparrow}U^\dagger} \\
&= \frac{1}{D^2d^L} \sum_{s_0, s_1, ..., s_{L+1}} 
\bbE{U\ket{S_0,S_1,...,S_{L+1}} \bra{S_0,S_1,...,S_{L+1}} U^\dagger} \\
&= \frac{1}{D^2d^L}\bbE{UU^\dagger}.
\label{Eq:B1}
\end{split}
\end{equation}
Note that $U_i$ is a unitary matrix, and we have $U_i U_i^\dagger = I$
[see Fig.~\ref{Fig:Fig7_Proof_P}(b)]. Recursively using the canonical
condition, we arrive at $\bbE{U\ket{\uparrow,\uparrow,...,\uparrow}
\bra{\uparrow,\uparrow,...,\uparrow}U^\dagger} = \frac{1}{Dd^L} I$. 

\section{Probability density in energy space}
\label{App:C}
Notice that the energy functional  
\begin{equation}
	E[\psi] := \frac{\expectationvalue{e^{-\beta H}H}{\psi}}{\expectationvalue{e^{-\beta H}}{\psi}}
\end{equation}
can be regarded as a random variable, which makes the energy space $\mathbb{R}$ a probability space with probability density 
\begin{equation}
	P(\epsilon) := \lim_{d\epsilon \rightarrow 0} \frac{\mathbb{P}(E \in [\epsilon, \epsilon + d\epsilon])}{d\epsilon}.
	\label{Eq:P}
\end{equation} 
where $\mathbb{P}$ denotes the probability measure in sample space. $P(\epsilon)$ represents the energy distribution of the samples and thus can be used to characterize the energy typicality, i.e. if the obsevrved energy of sample states always has a high probability to be close to the average energy or not.  

Note the average energy is not the expectation value corresponding to $P(\epsilon)$, instead, a Boltzmann weight $W[\psi] := \expectationvalue{e^{-\beta H}}{\psi}$ is needed, i.e.
\begin{equation}
	\begin{split}
		E(\beta) &=\frac{1}{Z(\beta)}\int E[\psi]W[\psi] \dP\\
		&= \frac{1}{Z(\beta)}\int d\epsilon \int \epsilon W[\psi] 
		\frac{\dP}{d\epsilon}\\
		&:= \int \epsilon P_W(\epsilon)d\epsilon.
	\end{split}
\end{equation}
Thus the weighted effective probability density reads
\begin{equation}
	P_W(\epsilon) = \frac{1}{Z(\beta)}\lim_{d\epsilon \rightarrow 0}\int\limits_{S_\epsilon^{d\epsilon}}
	W[\psi]\frac{\dP}{d\epsilon},
	\label{Eq:PW}
\end{equation} 
where $S_\epsilon^{d\epsilon} = \{\ket{\psi}| E[\psi] \in [\epsilon, \epsilon + d\epsilon]\}$ denotes the $[\epsilon, \epsilon + {\rm d}\epsilon]$ energy shell in sample space. 

Both Eq.~(\ref{Eq:P}) and Eq.~(\ref{Eq:PW}) can
be estimated via standard kernel density estimation methods, resulting in Eq.~(\ref{Eq:P_e}) and Eq.~(\ref{Eq:PW_e}) in the main text, respectively.

\bibliography{stoMPSRef}

\end{document}